\documentclass{article}
\usepackage{spconf,amsmath,graphicx}
\usepackage{amssymb}
\usepackage{bm}
\usepackage{caption}

\usepackage[sort&compress,numbers]{natbib}
\usepackage{siunitx} 
\DeclareSIUnit{\pixel}{px}
\usepackage{booktabs}
\usepackage{xspace}
\makeatletter
\DeclareRobustCommand\onedot{\futurelet\@let@token\@onedot}
\def\@onedot{\ifx\@let@token.\else.\null\fi\xspace}
\def\eg{e.g\onedot}

\makeatother

\renewcommand{\geq}{\geqslant}

\usepackage{textalpha}

\usepackage{booktabs}
\usepackage{makecell}
\usepackage{graphicx}
\usepackage{microtype}

\usepackage{enumitem}
\setlist{nosep}

\usepackage{textcomp}

\newcommand*{\tinyskip}{\par\vspace{1.5pt plus 0.5pt minus 0.5pt}}

\title{Beyond the section plane: Robust 3D reconstruction of multiplex immunofluorescence exposes limitations of 2D spatial cell statistics}
\name{\begin{tabular}{c}%
Junming Ma$^{\kern1pt\text{1}}$, 
Olle Edgren Sch\"{u}llerqvist$^{\kern1pt\text{1,2}}$, 
Hui Yu$^{\kern1pt\text{3}}$, 
Carina Strell$^{\kern1pt\text{3}}$, 
Patrick Micke$^{\kern1pt\text{3}}$, \\
Joakim Lindblad$^{\kern1pt\text{1,2}}$, 
Artur Mezheyeuski$^{\kern1pt\text{4}}$, 
Nata\v{s}a Sladoje$^{\kern1pt\text{1,2}}$
\end{tabular}}
\address{Author Affiliation(s)}
\address{%
$^{1}$ Department of Information Technology, Uppsala University, Uppsala, Sweden \\
$^{2}$ SciLifeLab, Uppsala University, Uppsala, Sweden \\ 
$^{3}$ Department of Immunology, Genetics and Pathology, Uppsala University, Uppsala, Sweden \\
$^{4}$ Vall d'Hebron Institute of Research \& Institute of Oncology,  Barcelona, Spain 
}%
\begin{document}
%
\maketitle
\begin{abstract}
%
Spatial analysis of the tumour microenvironment with multiplex immunofluorescence (mIF) typically relies on single thin tissue sections, which misrepresent the 3D tissue organisation and bias spatial cell statistics in ways that remain largely unquantified. 
We reconstruct 24 breast and lung tumour tissue microarray cores in 3D, each from 15--20 serial mIF sections, and compare spatial cell statistics 
in 3D with those from individual sections.
Our 
pipeline combines affine pre-alignment
with dense deformable refinement using RoMa~v2, achieving a median target registration error of 1.79 µm, well below nuclear size, which allows nuclei to be  
linked across sections. 
While cell-type proportions are preserved in 2D (Pearson corr. $r\geq0.99$), 3D analysis reveals up to 3.8 times shorter nearest-neighbour distances, 13--19\% higher neighbourhood entropy, and 2--4 times more contacts per cell. 
Tissue composition can thus be reliably estimated 
from individual sections, whereas 
cell-cell interactions
require volumetric analysis. 



\end{abstract}
\section{Introduction and related work}
\label{sec:intro}

The tumour microenvironment (TME) is a 3D ecosystem of cancer, immune, and stromal cells, vessels, and extracellular matrix, whose spatial organisation is linked to patient prognosis and response to therapy~\cite{GALON201311, Lu2019}. Highly multiplex imaging, such as multiplex immunofluorescence (mIF) and imaging mass cytometry (IMC), measures tens of protein markers per cell while preserving tissue architecture, facilitating spatial single-cell analysis as a central tool in cancer research~\cite{Semba2024, sorin2023, mezheyeuski2023}. Despite the 3D nature of the TME, nearly all mIF studies analyse individual \qtyrange{3}{5}{\micro\meter} thick sections. Volumetric alternatives, such as light-sheet microscopy with tissue clearing or 3D IMC~\cite{Kuett2022}, are costly and slow~\cite{cho2023}.

Single-section analysis implicitly assumes that one thin section represents the tissue. As cells are comparable in size to the section thickness,
2D estimates of their counts, distances, contacts, and neighbourhoods are geometrically biased. 
Stereological corrections~\cite{gundersen1988} address cell counts and sizes, but 
not relational metrics such as nearest-neighbour distances 
or neighbourhood composition. 
3D imaging shows that spatial features vary substantially between nearby sections, 
and that incomplete information causes 
inaccurate phenotyping and many obscured cell--cell contacts~\cite{yapp2025}.

3D reconstruction from serial sections offers an
alternative compatible with routine workflows, but requires registering sections despite
tears and 
deformations from cutting and
mounting~\cite{kurz2024}. 
Rigid and non-rigid approaches are often sequentially applied. 
VALIS~\cite{Gatenbee2023}
performs rigid alignment using modern feature detectors and matchers (\eg, DISK, DeDoDe and LightGlueMatcher),
followed by SimpleElastix or RAFT for non-rigid adjustments. 
Ghose et al.~\cite{Ghose2023} limit drift along the stack by calibrating mIF alignment against micro-CT, at the cost of additional imaging.
For H\&E histology, CODA~\cite{CODA2022} cross-correlates Radon transforms of single-channel images, while RAPID~\cite{RAPID2026} 
uses dense feature matching by RoMa~\cite{RoMa}. 
The recent RoMa~v2~\cite{romav2} achieves state-of-the-art accuracy and efficiency in dense feature matching but, to our knowledge, has not been evaluated on histopathology data. 

 mIF of clinical formalin-fixed paraffin-embedded (FFPE) material adds further  challenges: short, irregularly spaced stacks,  damaged or missing sections, large deformations, inter-section staining variability, and repetitive nuclear texture that confounds sparse matching. Moreover, single-cell spatial statistics demand cell-level registration accuracy, whereas existing tools are typically validated at the tissue level. 
 

\tinyskip\noindent\textbf{Our contributions are:}
\begin{enumerate}[label=(\roman*),leftmargin=18pt] 
\item a deformable serial-section registration pipeline built on RoMa~v2 for accurate 3D reconstruction of mIF;
\item evaluation at a cellular scale based on expert-annotated landmarks against 
established baselines 
(affine, VALIS); 
\item a 3D cell reconstruction method, linking nuclei across aligned sections; 
\item 
quantification of a bias of single-section TME analysis, via comparison of cell-type proportions, nearest-neighbour distances, neighbourhood composition, and cell contacts, in 2D and 3D. 
\end{enumerate}

\tinyskip\noindent
We publish the code and detailed parameters of the pipeline at https://github.com/MIDA-group/beyond-section-plane. 


\section{Data}
\label{sec:data}

We analyse an in-house tissue microarray (TMA) dataset prepared by the Micke-Strell lab at Uppsala University. It consists of 30 FFPE patient cores 
(\qty{2}{\milli\meter} diameter), split evenly between breast and lung tumour tissue. 
%
Each core comprises 20 serial \qty{4}{\micro\meter} sections, stained with a seven-colour Opal panel and imaged on a PhenoImager HT (Akoya Biosciences) at a pixel size of \qty{0.4961}{\micro\meter}. After spectral unmixing, each section has eight channels: DAPI (nuclei), CD31 (endothelial cells), GAP43 and NFP (nerve-associated cells), CD3 (T cells), CD163 (M2-like macrophages), PanCK (epithelial/tumour cells), and autofluorescence.

\tinyskip\noindent\textbf{Data Curation.}
%
From the acquired stack of 20 images, each including slices of 30 tissue cores, 6 cores were excluded due to observed severe tissue damage across all slices. To ensure meaningful downstream processing, we further eliminate a few heavily damaged individual sections within the remaining cores. This yields a final dataset of 24 cores, each containing between 15 and 20 consecutive tissue sections.

\section{Methods}
\label{sec:methods}


\subsection{Serial-section Registration}
\noindent\textbf{Preprocessing.} 
In each section, cores are detected, rotationally corrected, and cropped. Tissue masks are computed from the normalised sum of all eight channels after background removal by morphological opening: a Triangle threshold of the smoothed image is intersected with a coarser Otsu foreground and cleaned by morphological closing and opening.

\tinyskip\noindent\textbf{Two-step registration.} 
The sections of each core are affinely registered 
from AKAZE keypoints~\cite{lester2017}, followed by dense non-rigid refinement with RoMa~v2. Starting from the
central section, registration proceeds outward in both directions,
aligning each section to its already registered neighbour. Tissue
masks undergo the same transformations, and restrict keypoint
detection, quality assessment, and warping to tissue.
%

\tinyskip\noindent\textbf{Affine alignment.} 
AKAZE keypoints are detected (threshold $10^{-4}$) within the tissue mask on the log-stretched PanCK channel and matched using Lowe's ratio test (threshold 0.8). The transformation is estimated with MAGSAC~\cite{MAGSAC} and refined by least squares on the inliers. This leaves only small residual displacements for the deformable stage, which can therefore be tightly bounded.

\tinyskip\noindent\textbf{Dense refinement.} Since RoMa~v2 
is designed for RGB data, we fuse the mIF channels of each section into a single RGB image. 
Best performance in our
testing of different channel combinations and preprocessing techniques was reached by tinting and summing the DAPI and PanCK channels enhanced by CLAHE~\cite{CLAHE}. 
RoMa~v2 generates a displacement grid 
at resolution $1280{\times}1280$ 
with 
a confidence value 
per grid point; 
background and low-confidence points (${<\,}0.75$) are assigned zero displacement, displacements are capped at 80 pixels (\qty{40}{\micro\meter}), and the grid is upsampled to full resolution. 

\tinyskip\noindent\textbf{Quality control.} Each stage is retained only if it improves alignment, measured by masked normalised cross-correlation (NCC) on the log-stretched PanCK channel. The affine transformation is replaced by the identity if it reduces NCC relative to the unregistered pair. 

\subsection{3D Cell Reconstruction}


Due to strong anisotropy (voxel
size $0.496{\times}0.496{\times}\qty{4.0}{\micro\meter}$) of the data, direct 3D cell segmentation with Cellpose v4.0.8 
performed poorly. 
Instead, we use Cellpose v3.1.1.2~\cite{Cellpose}  with the cyto3  model 
to segment nuclei in 2D on the unregistered DAPI channel, and link them across sections using denoised markers. 

\tinyskip\noindent\textbf{Marker denoising.} 
Images are denoised with a $3{\times}3$ median filter. Artefacts are detected via channel-specific rules (morphological for DAPI; difference-of-Gaussians otherwise), inpainted, and zeroed. Backgrounds are subtracted using morphological top-hat filters (radius \qty{8}{\micro\meter} for DAPI/CD3/CD163; \qty{40}{\micro\meter} for PanCK) or broad Gaussian blurs ($\sigma = \qty{80}{\micro\meter}$ for CD31; \qty{60}{\micro\meter} for GAP43/NFP).

\tinyskip\noindent\textbf{Linking nuclei across sections.} 
With \qty{4}{\micro\meter} thick sections, a single nucleus may appear in multiple consecutive slices. 
To assemble the 3D representation we construct a graph whose nodes are 2D nuclear masks and whose edges connect masks in adjacent sections overlapping by at least 30 pixels; edges are weighted by the Intersection-over-Union (IoU) of the connected pair. 
Since nuclei are expected to change shape only gradually between sections, 
edges between elongated masks (eccentricity $\geq 0.5$) 
whose major axes differ by more than \ang{35} are removed
as indicating distinct overlapping nuclei. 
Each mask retains only its highest-scoring link to each neighbouring section
(IoU, linearly down-weighted by orientation difference). 
The connected
components of the resulting graph form candidate 3D nuclei. Components spanning more than five sections are recursively split at their lowest-scoring edge, and
components 
with a largest 2D mask below
100 pixels are discarded. 

\tinyskip\noindent\textbf{Cell typing.} 
Each 2D nuclear mask is dilated by 4 pixels (\qty{2}{\micro\meter}) to capture phenotypic markers in the membrane and cytoplasm,
and the median marker intensity within this region is tested against
an ordered rule list where the first match applies:
T cells (CD3$^+$), Nerve-associated cells (GAP43$^+$ or NFP$^+$), Epithelial/Tumor cells (PanCK$^+$), 
Endothelial (CD31$^+$),
Macrophage (CD163$^+$), otherwise Unknown. 

Marker thresholds are core-wide medians of per-section values, computed using Otsu's method (GAP43, NFP, PanCK), a fixed threshold of 200 (CD31), or a bimodal Gaussian mixture model (CD3, CD163; Otsu's fallback if $\Delta\mathrm{BIC}{\,\le\,}6$). 
Cells are marked PanCK-positive if ${\ge\,}50\%$ of their expanded nuclear mask overlaps a morphologically refined PanCK mask derived from the PanCK threshold. Final 3D cell types are assigned via area-weighted majority vote of positive 2D calls.


\subsection{Statistical Characterisation of the TME}
We compare four established metrics for TME characterisation ~\cite{BACKMAN2023,Ghose2023,Kuett2022}.
2D metrics are computed per section and 3D metrics on the reconstructed volume.
 
 
\tinyskip\noindent\textbf{Cell-type proportions.}
Cell-type proportions are expressed as the relative densities of the five specific typed classes. In 2D, density is the nucleus count divided by the convex-hull area of the slice centroids, computed per slice and averaged across slices. In 3D, it is the total reconstructed nucleus count divided by the convex-hull volume, giving a single value per core. The density of each cell type is normalised by the total density of these five types.

\tinyskip\noindent\textbf{Nearest-neighbour distances.} 
Euclidean distances from each source cell to the nearest target cell (excluding itself for same-type pairs) are computed within sections (2D) or across the reconstructed volume (3D). Distances are averaged over source cells per type pair; 2D values are averaged per section, then across sections to yield a single 2D value per core.
%

\tinyskip\noindent\textbf{Neighbourhood composition.} 
The Shannon entropy of cell types among all neighbours within \qty{30}{\micro\meter} of each cell is computed within sections (2D, a disc) or across the reconstructed volume (3D, a sphere). To avoid edge effects, only cells in the central \qty{12}{\micro\meter} of the stack are evaluated: the three middle sections in 2D and cells with centres within \textpm\qty{6}{\micro\meter} of the volume's central z-plane in 3D. Entropies are averaged over cells per type. 

%

\tinyskip\noindent\textbf{Cell contacts.} 
Nuclei are dilated by \qty{2}{\micro\meter}, and contacts between dilated cells are identified within sections (2D) or across the reconstructed volume (3D). The number of distinct contacting target cells is summed per type pair and divided by the number of source cells, including those without contacts.

%
%

\section{Experiments and Results}
\label{sec:experiments}

\subsection{Registration Evaluation}

Registration accuracy is measured by target registration error (TRE), the Euclidean distance between corresponding landmarks after registration, on 284 landmark pairs placed by a domain expert in adjacent sections of four cores. Since landmarks are placed manually on different cross-sections of the same nucleus, TRE is not expected to reach zero even under perfect alignment.

We compare our full pipeline 
with its affine stage alone and with VALIS~\cite{Gatenbee2023}. 
For this comparison we use VALIS v1.2.0 with default settings; adjacent sections are registered using VGG-features with 
VALIS standard matcher, followed by 
alignment using DISK features matched with LightGlue. Correspondences are filtered with MAGSAC to estimate a similarity transform between adjacent sections, followed by non-rigid registration with VALIS optical-flow warper.

Table~\ref{tab:tre_roma_valis} summarises the 
results. The dense refinement reduces the median TRE from 3.62 (
by affine only 
) to 1.79~\unit{\micro\meter} ($\sim$3.5 pixels), approaching the limit of manual annotation accuracy. Our pipeline and VALIS achieve a similar median TRE (1.79 vs.\ 1.81~\unit{\micro\meter}), but ours yields lower mean, maximum, and 90th-percentile errors (4.27 vs.\ 5.49~\unit{\micro\meter}), 
keeping 90\% of errors well below a typical nuclear diameter, indicating fewer large registration failures and confirming the sub-cellular accuracy required for single-cell spatial statistics.

\begin{table} 
\caption{Target registration error (TRE) over 284 
expert-annotated landmark pairs across four annotated cores: before registration (Raw), after affine alignment, after both affine and deformable alignment (Our) and using VALIS~\cite{Gatenbee2023}. 
}
%
\label{tab:tre_roma_valis}
\centering
\small 
\begin{tabular}{lrrrr}
\midrule
Metric & \textbf{Raw} & \textbf{Affine} & \textbf{Our} & \textbf{VALIS} \\
\midrule
Mean TRE ($\mu$m) & 25.03 & 4.86 & \textbf{2.59} & 3.03 \\
Std TRE ($\mu$m) & 28.86 & 3.81 & \textbf{2.79} & 4.60 \\
Median TRE ($\mu$m) & 20.89 & 3.62 & \textbf{1.79} & 1.81 \\
P90 TRE ($\mu$m) & 39.71 & 8.99 & \textbf{4.27} & 5.49 \\
Max TRE ($\mu$m) & 335.98 & 26.43 & \textbf{22.91} & 35.62 \\
\bottomrule
\end{tabular}
\end{table}

\begin{figure*}[t!]
    \centering
    \includegraphics[width=\textwidth]{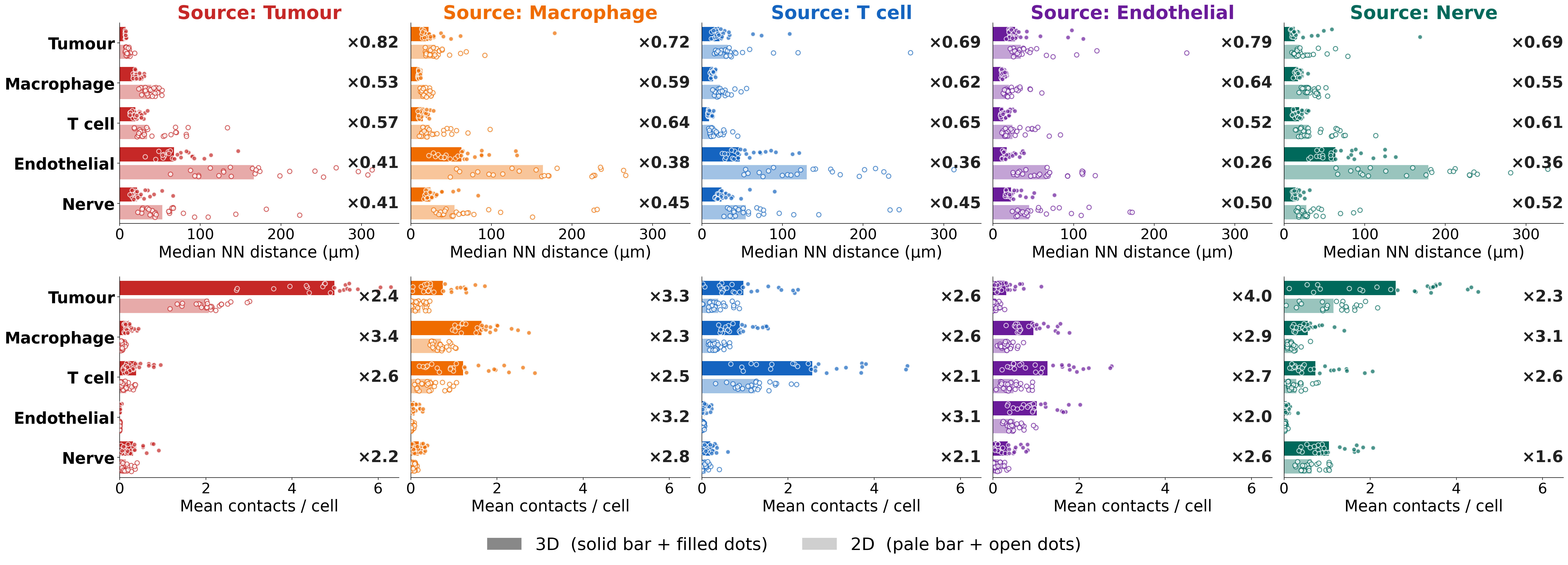}
    \captionsetup{margin=2cm}
    \caption{Median nearest-neighbour distances (top) and mean contacts per cell (bottom) in 3D vs.\ 2D for each source (columns) and target (rows) cell type across 24 cores.} 
    \label{fig:nn_distances_contacts}
\end{figure*}

\subsection{3D Cell Reconstruction}
Across 24 cores, we segment \num{4787548} nuclei, linking 45.9\% across multiple sections. Linked nuclei span an average of 2.62 slices, aligning with the theoretical $\approx$2.75 for $\sim$\qty{7}{\micro\meter} nuclei at \qty{4}{\micro\meter} thickness. Among the 54.1\% unlinked single-section nuclei, the majority (30.1\%) exhibit sufficient spatial overlap ($\ge$30\,px) but fail our conservative linking criteria. The rest lack sufficient overlap (12.0\%), reside at stack boundaries (9.3\%), or remain unmatched (2.7\%). Consequently, the multi-section count represents a predicted lower bound for true multi-section cells.
Typing agreement, the fraction of multi-section cells whose voting slices all share the same type, decreases with depth: 81.3\%, 71.3\%, 63.7\%, and 58.8\% for cells spanning 2–5 sections.

\subsection{Statistical Characterisation of the TME in 2D \& 3D}


\begin{table}[!b]
\vspace*{-3mm}
\caption{
Cell-type proportions and relative neighbourhood entropy shift in 3D reconstructions vs.\ 2D sections across 24 cores. Cell-type proportions: mean$\pm$std of per-core fractions, $\Delta$ = 3D $-$ 2D, $r$ = Pearson correlation of 3D vs.\ 2D. Entropy shift: median [interquartile range] of per-core relative change; entropy 
is higher in 3D in every core for every cell type. 
}
\label{tab:proportions_entropy}
\centering
\small
\setlength{\tabcolsep}{3pt}
\renewcommand{\arraystretch}{1.1}
\resizebox{\columnwidth}{!}{%
\begin{tabular}{l r@{}l r@{}l r@{.}l @{} c c}
\toprule
& \multicolumn{7}{c}{\textbf{Cell-type proportions}} & \multicolumn{1}{c}{\textbf{Entropy shift}} \\
\cmidrule(lr){2-8} \cmidrule(lr){9-9}
Cell type & \multicolumn{2}{c}{3D (\%)} & \multicolumn{2}{c}{2D (\%)} & \multicolumn{2}{c}{$\Delta$ (pp)} & $r$ & Median [IQR] (\%) \\
\midrule
Tumour      & $50.5$ & $\scriptscriptstyle\pm19.2$ & $51.4$ & $\scriptscriptstyle\pm19.2$ & $-0$ & 9 & 1.00 & +18.7\;\;[14.4, 23.3] \\
Macro.      & $17.6$ & $\scriptscriptstyle\pm8.3$ & $16.5$ & $\scriptscriptstyle\pm8.3$ & $+1$ & 2 & 0.99 & +14.2\;\;[11.1, 17.7] \\
T cell      & $24.5$ & $\scriptscriptstyle\pm16.3$ & $24.5$ & $\scriptscriptstyle\pm16.7$ & 0    & 0 & 1.00 & +14.3\;\;[11.3, 16.3] \\
Endo.       & $1.3$  & $\scriptscriptstyle\pm1.0$ & $1.1$  & $\scriptscriptstyle\pm0.9$  & $+0$ & 2 & 0.99 & +12.6\;\;[10.0, 17.2] \\
Nerve       & $6.1$  & $\scriptscriptstyle\pm4.5$  & $6.4$  & $\scriptscriptstyle\pm4.9$  & $-0$ & 4 & 0.99 & +14.8\;\;[12.3, 20.6] \\
\bottomrule
\end{tabular}
}%
\end{table}

\tinyskip\noindent\textbf{Cell-type proportions.} 
Table~\ref{tab:proportions_entropy} compares cell-type proportions in the reconstructed 3D volumes with those estimated from 2D sections across all 24 cores. The two views agree closely: mean per-core differences reach at most 1.2 percentage points for all cell types, and 3D proportions follow the variation between cores captured in 2D almost perfectly ($r\geq0.99$). This holds for both dominant cell populations (tumour cells) and minor ones (endothelial and nerve cells). 


\tinyskip\noindent\textbf{Nearest-neighbour distances.} 
Nearest-neighbour distances are shorter in 3D than in 2D for every source-target cell-type pair, as shown in Fig.~\ref{fig:nn_distances_contacts} (top). The factor depends on the cell type. Distances to tumour cells, the most abundant population, shrink least, followed by T cells and macrophages. Distances to endothelial and nerve-associated cells, the two least abundant populations, shrink most. The largest reduction occurs between endothelial cell-pairs (3D\,/\,2D ratio of 0.26).

\tinyskip\noindent\textbf{Neighbourhood composition.} 
Table~\ref{tab:proportions_entropy} summarises the relative neighbourhood entropy shift per cell type across all cores. Entropy is higher in 3D for each core and every cell type, indicating that 3D volumes capture systematically more diverse cellular microenvironments. Again, the differences are cell-type dependent, with median relative differences ranging from +12.6\% for endothelial cells to +18.7\% for tumour cells. 


\tinyskip\noindent\textbf{Cell contacts.} 
The number of cell contacts are consistently higher in 3D than in 2D, as shown in Fig.~\ref{fig:nn_distances_contacts} (bottom). Across cell-type pairs, 3D analysis yields 1.6 to 4.0 times more contacts per cell, with the largest absolute difference observed for tumour–tumour contacts.

\section{Discussion and Conclusion}
\label{sec:conclusion}

Our registration pipeline achieves errors well below the nuclear
diameter, enabling nuclei to be linked across sections and 3D spatial statistics to be extracted. 
Our findings show that while individual 2D sections reliably estimate cell-type proportions, they misrepresent spatial relationships in the microenvironment.


This discrepancy arises largely from cells outside the section plane, which volumetric analysis captures. Its magnitude depends on the structural arrangement of each cell type: shorter 3D distances between endothelial cells likely reflect vessel continuity across sections, while higher neighbourhood entropy and more homotypic contacts among tumour cells are consistent with dense structures spanning multiple sections. Because these biases depend on local geometry, they are unlikely to be resolved by global stereological corrections. Since they persist when averaging over sections, they are not remedied by additional 2D sampling. Spatial biomarkers derived from single sections should thus be interpreted with caution.


Two limitations remain. First, phenotype agreement decreases for cells spanning multiple sections, reflecting linking errors, batch effects, or biological marker variation. Second, the limited stack depth (60 to \qty{80}{\micro\meter}) truncates neighbourhoods, overestimating 3D distances to rare cell types; the reported 2D bias is thus likely conservative. Future work will use deeper volumes, validate against volumetric ground truth, and explore
correction models for single-section statistics.

\clearpage

\section*{Compliance with Ethical Standards}
\label{sec:ethics}
The study was conducted retrospectively using
human subject data. The analysis of the cohort was approved by the Regional Ethical Review Board in Uppsala, Sweden (2012/532).



\vspace{-1em}
\section*{Acknowledgments}
\label{sec:acknowledgments}
We report no conflicts of interests. This work is supported by Swedish Cancer Society projects 25 4871 Pj and 25 4859 Pj, the Swedish Research Council grant 2022-03580, and the SciLifeLab \& Wallenberg DDLS Program (grant KAW2024.0159). Computations were enabled by the Berzelius resource provided by the Knut and Alice Wallenberg Foundation at the National Supercomputer Centre.

\tinyskip\noindent\textbf{AI usage.} 
The authors used Claude (including Claude Opus 5.5 and Sonnet 4.6, Anthropic) for language editing, structural proofreading, condensing of author-written text, and generating boilerplate code snippets. The tools were not used to design the method, generate or analyse data, or produce results. All content was reviewed and edited by the authors, who take full responsibility for the publication.




\vspace{-1em}
\bibliographystyle{IEEEbib}
\bibliography{refs}


\end{document}